\documentclass[12pt,preprint]{aastex}

\newcommand{\kev}{keV}
\newcommand{\fe}{Fe~K$\alpha$}
\newcommand{\etal}{et al.}

\newcommand{\ixo}{\textit{IXO}}
\newcommand{\chandra}{\textit{Chandra}}
\newcommand{\xmm}{\textit{XMM-Newton}}

\shorttitle{The Integrated Relativistic \fe\ Line}
\shortauthors{Ballantyne}

\begin{document} 

\title{The Integrated Relativistic Iron Line from Active Galactic
  Nuclei: Chasing the Spin Evolution of Supermassive Black Holes}

%% Use \author, \affil, and the \and command to format
%% author and affiliation information.

\author{D. R. Ballantyne}
\affil{Center for Relativistic Astrophysics, School of Physics,
  Georgia Institute of Technology, Atlanta, GA 30332}
\email{david.ballantyne@physics.gatech.edu}

\begin{abstract}
The spin of a supermassive black hole (SMBH) is directly related to
the radiative efficiency of accretion on to the hole, and therefore
impacts the amount of fuel required for the black hole to reach a certain
mass. Thus, a knowledge of the SMBH spin distribution and evolution is
necessary to develop a comprehensive theory of the growth of SMBHs and
their impact on galaxy formation. Currently, the only direct
measurement of SMBH spin is through fitting the broad \fe\ line in
AGNs. The evolution of spins could be determined by fitting the broad
line in the integrated spectra of AGNs over different redshift
intervals. The accuracy of these measurements will depend on the
observed integrated line strength. Here, we present theoretical predictions of the integrated
relativistic \fe\ line strength as a function of redshift and AGN
luminosity. The equivalent widths of the integrated lines are much less
than $300$~eV. Searches for the integrated line will be easiest for
unobscured AGNs with $2$--$10$~\kev\ luminosities between $44 < \log L_{X} \leq
45$. The total integrated line makes up less
than $4$\% of the X-ray background, but its shape is sensitive to the
average SMBH spin. By following these recommendations, future
\textit{International X-ray Observatory} surveys of broad \fe\ lines
should be able to determine the spin evolution of SMBHs.
\end{abstract}


\section{Introduction}
\label{sect:intro}
In astrophysics, black holes can be described by only two parameters:
their mass $M_{\mathrm{BH}}$ and their angular momentum $J$
(parametrized by the dimensionless spin parameter $a \equiv
cJ/GM_{\mathrm{BH}}^2$ where $0 \leq a \leq 1$). Over the last several
decades, estimates of black hole masses have been obtained by
measuring the gravitational influence of the hole on nearby objects
such as binary stellar companions \citep[e.g.,][]{bol72,cow83,mr06},
orbiting stars \citep[e.g.,][]{ghez08}, stellar cusps
\citep[e.g.,][]{mag98} or dense clouds of ionized gas
\citep[e.g.,][]{peter04}. However, measurements of black hole spin are
much more difficult to obtain as the spin only significantly affects
the shape of space-time out to $\approx 6$~$r_g$
($r_g=GM_{\mathrm{BH}}/c^2$ is a gravitational radius) from the event
horizon \citep{bpt72}. In particular, the value of the spin is
directly related to the radius of the innermost stable circular orbit
(ISCO) around the hole \citep{bpt72}, and therefore impacts the
radiative efficiency of thin accretion disks. If the accretion flow is
rotating in the same sense as the hole, then a larger $a$ brings the
ISCO closer to the event horizon, increasing the efficiency that an
accretion flow can convert gravitational potential energy into
radiation. A larger fraction of the rest-mass energy is radiated away
in a highly efficient accretion disk, reducing the growth rate of the
black hole mass. Therefore, measuring the distribution of spins will
elucidate the role of steady accretion \citep[e.g.,][]{v05}, chaotic
accretion \citep[e.g.,][]{kp06} and black hole mergers
\citep[e.g.,][]{bv08} in the build-up of supermassive black holes
(SMBHs).

X-ray variability and optical microlensing data show that the X-ray
source in AGNs is within 10~$r_g$ \citep[e.g.,][]{iwa04,dai10},
indicating that the accretion disk is likely illuminated by X-rays
down to the ISCO. As spin clearly impacts the radius of the ISCO, the
extent of the red wing of the relativistically broadened \fe\ line is
a direct measurement of black hole spin \citep[e.g.,][]{rf08}. In
recent years, long observations by \xmm\ and \textit{Suzaku} have used
this method to obtain SMBH spin estimates for a small number of
bright, nearby Seyfert~1 galaxies \citep{br06,min09,sch09}. These type
of spin measurements depend on precise spectral fitting to the broad
\fe\ line and continuum and are most easily performed on sources with
strong \fe\ lines. Thus, the larger collecting area provided by the
future \textit{International X-ray Observatory} (\ixo) will be
necessary in order to extend these types of spin measurements over a significant
range of redshift. In the meantime, attempts have been made to measure the average \fe\ line profile by
stacking the X-ray data of the many AGNs detected in deep
\textit{Chandra} or \xmm\ surveys over a wide range of redshift and
AGN luminosity. If a significant \fe\ signal can be detected in the
integrated spectrum then, in principle, the average spin of the black
holes contributing to the line could be determined. Again, the ability
to make this measurement depends crucially on the strength of the
integrated \fe\ line. \citet{streb05} averaged the \xmm\ spectra of type 1
and type 2 AGNs detected in the Lockman Hole and found a strong
(equivalent width [EW] $\sim 500$~eV) and broad \fe\ line that
indicated a non-zero spin for the average SMBHs in their
sample. However, similar stacking analyses of different datasets by
\citet{bgc05} and \citet{cor08} were not able to detect a broad
component, with \citet{cor08} finding that the EW of the broad
component must be $< 400$~eV. These attempts to measure the integrated
broad \fe\ line are technically challenging and are limited by the
unknown intrinsic distribution of broad \fe\ line strengths. 

Recently, \citet{bal10} described a method by which the average EW of
the broad \fe\ line could be calculated as a function of $z$ and
luminosity. That paper concentrated on the distribution of likely EWs
from individual sources, and did not consider the result of
integrating lines over specific redshift intervals, a step crucial to
measuring the spin evolution of SMBHs. This letter builds on the
\citet{bal10} method to calculate the integrated \fe\ line over
several different redshift and luminosity ranges. We also consider the
dependence of AGN obscuration on the appearance of the integrated \fe\
lines. These predictions show which regions in the redshift and
AGN luminosity plane will have the largest average broad \fe\ EW, and
will be vital for guiding current and future attempts to map
out the spin evolution of SMBHs by fitting the average \fe\ line
profile. The calculations are reviewed in the next section, and the
results are presented in Sect.~\ref{sect:res}. The following
$\Lambda$-dominated cosmology is assumed in this paper:
$H_0=70$~km~s$^{-1}$~Mpc$^{-1}$, $\Omega_{\Lambda}=0.7$, and
$\Omega_{m}=0.3$ \citep{spe03}.

\section{Calculations}
\label{sect:calc}
The calculation of the mean \fe\ EW as a function of $z$ and AGN
$2$--$10$~\kev\ luminosity, $L_{X}$, broadly follows the procedure
described by \citet{bal10}. The \citet{ueda03} hard X-ray luminosity
function (HXLF) is combined with the observed AGN black hole mass
function (BHMF) at $z=0.15$ \citep{net09} to calculate an Eddington ratio
distribution at a given $L_{X}$. In order to predict the integrated
\fe\ line over several cosmologically interesting redshift ranges we assume that the BHMF evolves as $(1+z)^{1.64}$ between $z=0$ and $5$ \citep{lab09}. It is
also assumed that the broad \fe\ line is produced by a single strong reflection event
from a dense, geometrically thin accretion disk that can be
approximated by a constant density slab \citep{rf93,bal01}. The EW of the iron
line depends on the photon index of the illuminating power-law
$\Gamma$, the relative abundance of iron $A_{\mathrm{Fe}}$, the
ionization parameter of the disk $\xi$, and the reflection fraction,
$R$, which determines the relative strength of the reflected spectrum in the total
observed spectrum \citep{bfr02}. The first three of
these parameters are known to depend on the Eddington ratio of the AGN
\citep{nt07,ith07,rye09}. Reflection spectra are then computed for each Eddington ratio
contributing to an observed $L_{X}$. The \fe\ EW is measured for each
spectrum (assuming $R=0.5$, $1$, $2$ or $4$) and
averaged over the Eddington ratio distribution to give a single value
for that $L_{X}$. In order for the integrated line to reflect the
changes in $\xi$ and $\Gamma$, we keep track of the average energy
at which the \fe\ line reaches a maximum $\left <E_{\mathrm{max}}
\right >$. These calculations are performed for 100 values of
$L_{X}$ between $\log L_{X}=41.5$ and $\log L_{X}=48$, and for 100
values of $z$ between $z=0$ and $5$.

These results can now be included in a X-ray background (XRB) synthesis
calculation to predict the strength and profile of the integrated
broad \fe\ line over several different ranges of luminosity and
redshift. The synthesis model used here was last described by \citet{db09},
although we neglect the contribution of blazars for this
application. The integrated rest-frame spectral intensity $I(E)$ are
computed by evaluating:
\begin{equation}
I(E) = {c \over H_0} \int_{z_{\mathrm{min}}}^{z_{\mathrm{max}}}
  \int_{\log L_{X}^{\mathrm{min}}}^{\log L_{X}^{\mathrm{max}}}
  {d\Phi(L_X,z) \over d\log L_X} {S_{E}(L_X,z) d_l^2 \over (1+z)^2
  (\Omega_m (1+z)^3 + \Omega_{\Lambda})^{1/2}} d\log L_X dz,
\label{eq:cxrb}
\end{equation} 
where $d\Phi(L_X,z)/d\log L_X$ is the \citet{ueda03} HXLF,
$S_{E}(L_X,z)$ is the absorbed rest-frame spectrum of an AGN with
intrinsic luminosity $L_X$ at redshift $z$, and $d_l$ is the
luminosity distance to redshift $z$. At each ($L_X,z$) pair in the
integral, a relativistically broadened \fe\ line appropriate for a
maximally-spinning (i.e., $a \sim 1$) SMBH is added to the
spectrum $S_{E}$ such that it has the EW predicted by the reflection
calculations described above. The relativistic profile for the line
was calculated using the `laor2' model in XSPEC12 \citep{lao91,arn96} with a line
emissivity of $0$ between $1.2$ and $6$~$r_g$ and $-3$ from $6$ to
$400$~$r_g$ \citep[cf.,][]{nan07}. The inclination angle of the line
emitting material was taken to be $30$~degrees, and the rest energy of the line is set to
$\left <E_{\mathrm{max}} \right >$ to account for changes in the average ionization
state of the line as $L_{X}$ and $z$ are varied. Finally, to be
consistent with the calculation of the \fe\ lines, the photon index of
$S_{E}(L_X,z)$ is derived from averaging $\Gamma$ over the Eddington
ratio distribution for a given ($L_X,z$) pair. This method translates
to average photon indices varying from $1.4$ to $2.5$ depending on the
value of $z$ or $L_{X}$. In practice, this method of determining
$\Gamma$ for $S_{E}(L_X,z)$ has a negligible effect on the results.

The integrated AGN spectra are then computed for five different
redshift ranges: $0 \leq z \leq 5$, $0 \leq z \leq 1$, $1 < z \leq 2$, $2 <
z \leq 3$ and $3 < z \leq 5$. For each redshift range, five intervals
in luminosity are considered: $41.5 \leq \log L_{X} \leq 43$, $43 < \log
L_{X} \leq 44$, $44 < \log L_{X} \leq 45$ and $45 < \log L_{X} \leq
48$. The EW of the \fe\ line is measured by directly
integrating the resulting spectra. The EW measurements are performed separately for type 1 AGNs (those with
hydrogen column densities $\log N_{\mathrm{H}} < 22$) and type 2 AGNs
(those with $22 \leq \log N_{\mathrm{H}} \leq 25$), although the EW of
the broad line in this model is independent of the obscuration toward
the AGN. The impact of obscuration will be discussed at the end of
Sect.~\ref{sub:ews}.

\section{Results}
\label{sect:res}
\subsection{Integrated EWs}
\label{sub:ews}
Figure~\ref{fig:ews} plots the measured EWs from the integrated spectra
described in the previous section. The solid lines
show the results for $R=1$ while the dashed and dotted lines assume
$R=2$ and $0.5$, respectively. The
colors denote the different redshift ranges over which the integration
was performed. The measured EWs in all cases are significantly less than $400$~eV,
consistent with the upper-limit derived by \citet{cor08}. Indeed, for $R=1$,
the EWs are between $\sim 90$~eV and
$\sim 220$~eV. In fact, even if $R=2$ for every AGN, the EW of the
integrated \fe\ line does not break 320~eV. If the average $R$ of AGNs
is closer to $0.5$ then the majority of the integrated broad lines
have EWs$< 120$~eV.

In the model of \fe\ production described here and by \citet{bal10}
the EW of the line is ultimately a function of the Eddington ratio of
the AGN. If the SMBH is accreting too weakly, there is no optically
thick disk for reflection to occur \citep{ny95}. In the opposite
extreme, if the accretion ratio is very close to the Eddington limit
there is also little observed reflection signature because the disk is
very highly ionized \citep{bfr02}. The strongest \fe\ line arises from
recombination onto He-like iron, but this occurs over a relatively
narrow range of ionization parameter \citep{bfr02}, and, therefore, a
correspondingly small range of Eddington ratios \citep{bal10}. This
interplay between Eddington ratios and \fe\ EWs is reflected in the
redshift distributions of the integrated EW seen in
Figure~\ref{fig:ews}. For all AGNs with $\log L_{X} < 45$ the largest
integrated EW is found from those with $z < 1$. As $z$ increases, the
average mass of active SMBHs increases, which, at a constant $L_{X}$,
corresponding to a decreasing Eddington ratio. Therefore, the
integrated \fe\ EW drops as $z$ increases because a larger fraction of
AGNs at these luminosities are accreting too weakly to produce a
strong disk line. Integrating over the entire redshift range (black
lines in Fig.~\ref{fig:ews}) results in a weighted average of this
behavior. In contrast, the \fe\ EWs of AGNs with $\log L_{X} > 45$ are
largest for $1 < z \leq 2$ and smallest for $z < 1$. The luminosities
of these AGNs are large enough that ionized \fe\ lines contribute
significantly at $z > 1$ and boost the integrated EW. In this case,
the Eddington ratios at lower $z$ are, on average, too large and the
resulting disk lines are weaker. At larger redshifts, the increasing
SMBH mass reduces the average Eddington ratio which results in strong
neutral \fe\ emission. This discussion illustrates that, for a fixed
luminosity range, the integrated line probes smaller Eddington ratios
as the redshift increases. Elucidating the spin evolution of the SMBH
population over a fixed interval of mass will require measuring the
integrated \fe\ line over ranges of progressively higher luminosity as
the redshift increases.

At face value, the results of Fig.~\ref{fig:ews} indicate that
experiments using the broad \fe\ line to measure the spin
evolution of SMBHs should be targeted toward integrating the spectra
of AGNs with $\log L_{X} > 45$, as these sources are the ones
predicted to have the strongest integrated EWs. The difficulty with this strategy is
that such AGNs are rare at all redshifts \citep[e.g.,][]{ueda03}, and
thus the compilation of suitable samples will be extremely problematic
even with the enhanced sensitivity of \ixo. Therefore, we conclude
that AGNs with $44 < \log L_{X} \leq 45$ will provide the best sample
to search for the integrated broad \fe\ line. These AGNs are close to
the knee of the HXLF at all redshifts \citep[e.g.,][]{ueda03} and will
therefore be common enough to provide a useful sample at several
different redshift bins. The broad lines from these AGNs
will show a mixture of of neutral and ionized reflection with
$\left <E_{\mathrm{max}} \right > \approx 6.5$--$6.6$~\kev.  

Examples of the integrated $R=1$ spectra in the $44 < \log L_{X}
\leq 45$ range are shown as solid lines in
Figure~\ref{fig:lines} with the colors denoting the redshift ranges as
in fig.~\ref{fig:ews}. These spectra were calculated assuming that the
\fe\ line was being emitted all the way down to the ISCO of a
maximally spinning black hole. The dashed lines show the integrated
spectra if the ISCO is at $6$~$r_g$, as would be the case for a
non-spinning black hole. As expected, this scenario causes the
integrated line to be slightly less broad and more peaked. This small 
difference in the line profile should be measurable if a large enough sample of AGNs
could be included in the integral. This result nicely illustrates the
importance of targeting AGN in the luminosity range that will produce
the most intense integrated broad \fe\ line.

The integrated spectra shown in figure~\ref{fig:lines} and
the associated EWs plotted in fig.~\ref{fig:ews} are compiled from
only type 1 AGNs. The dotted blue line in Fig.~\ref{fig:lines} plots
the integrated type 2 spectrum with $44 < \log L_{X} \leq 45$ and $z \leq 1$. Our calculations assume that there is no difference in the
broad \fe\ EW between the type 1 and 2 AGNs. However, a visual
inspection of the type 2 spectrum in fig.~\ref{fig:lines} seems to
indicate a much more intense broad \fe\ line. Indeed, performing the same
integration of the type 2 line gives an EW of $460$~eV as
compared to $180$~eV for the type 1 AGN. The strong absorption in the
average type 2 spectrum causes the continuum to curve downwards at
energies $< 5$~eV and enhances the Fe K edge at $7.1$~\kev. These
effects conspire to artificially enhance the EW of the broad \fe\
line. Therefore, unless the underlying spectrum and absorption
distribution is well known, the search for broad \fe\ lines in
integrated spectra should only be performed on unobscured, type 1
AGNs.

\subsection{The Broad Line in the X-ray Background}
\label{sub:feback} 
\begin{figure}
\includegraphics[width=0.95\textwidth]{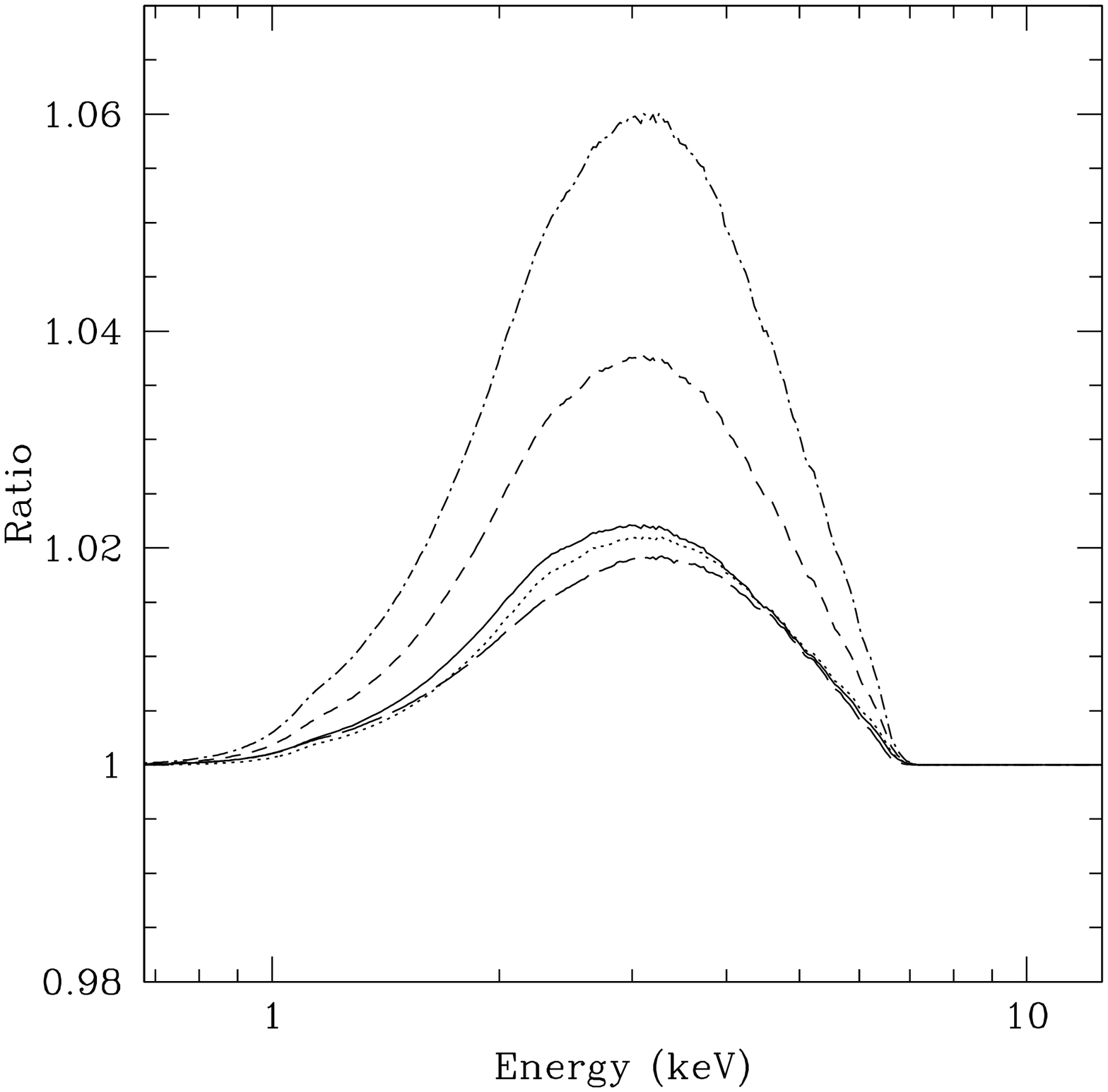}
\caption{The ratio of a model XRB spectrum that includes the broad \fe\ line
  to one that does not include the broad line. The ratios shown
  include a model with $R=1$ (solid line), $R=2$ (short-dashed line) and
  $R=4$ (dot-dashed line). Also shown is the ratio from a model where
  each AGN has a neutral broad \fe\ with a constant EW of 100~eV
  (long-dashed line). Finally, the dotted line corresponds to a $R=1$
  model where all the SMBHs are non-spinning and the lines extend to
  only $6$~$r_g$.}
\label{fig:ratio}
\end{figure}

Performing the integral of eqn.~\ref{eq:cxrb} over all $z$ and $\log
L_{X}$, but now redshifting $S_{E}$ into the observed frame results in
a prediction of the contribution of the \fe\ line to the entire XRB. Other authors have made similar predictions \citep{gil99,gf03}, but have had
to make very simple assumptions on the \fe\ strength and
shape. Figure~\ref{fig:ratio} shows the first prediction of the \fe\
contribution to the XRB that arises from a physical model
of the \fe\ line distribution. This figure plots the ratio of the XRB
spectrum including the broad line to one which does not include the
broad line. Predictions are shown for $R=1$ (solid line), $2$
(short-dashed line) and $4$ (dot-dashed line). The long-dashed line
plots the ratio for a model where the broad line has a constant EW of
100~eV and results from neutral reflection. In all of these cases the
line is broadening using the `laor2' model as described in
Sect.~\ref{sect:calc}, except for the dotted line which plots the
$R=1$ model with all lines arising from $6$~$r_g$ (also using the `laor' model).

Figure~\ref{fig:ratio} shows that the maximum contribution of the broad
\fe\ line to the XRB is $6$\%, but, since intense reflection cannot
be a common occurrence in AGNs \citep{gan07,bal10}, the contribution is more
realistically much less than $4$\%. These values are in agreement with
the previous predictions of \citet{gil99} and \citet{gf03}. The contribution of
integrating multiple observed-frame spectra and relativistic blurring
smears the \fe\ contribution over 6~\kev\ in energy, but there is a
slight difference between the profiles with different spin
distributions. Therefore, given a very high quality spectrum of the
X-ray background, the total integrated \fe\ profile will be able to
determine the average spin of SMBHs over cosmic time.

\section{Conclusions}
\label{concl}
\begin{figure}
\includegraphics[width=0.95\textwidth]{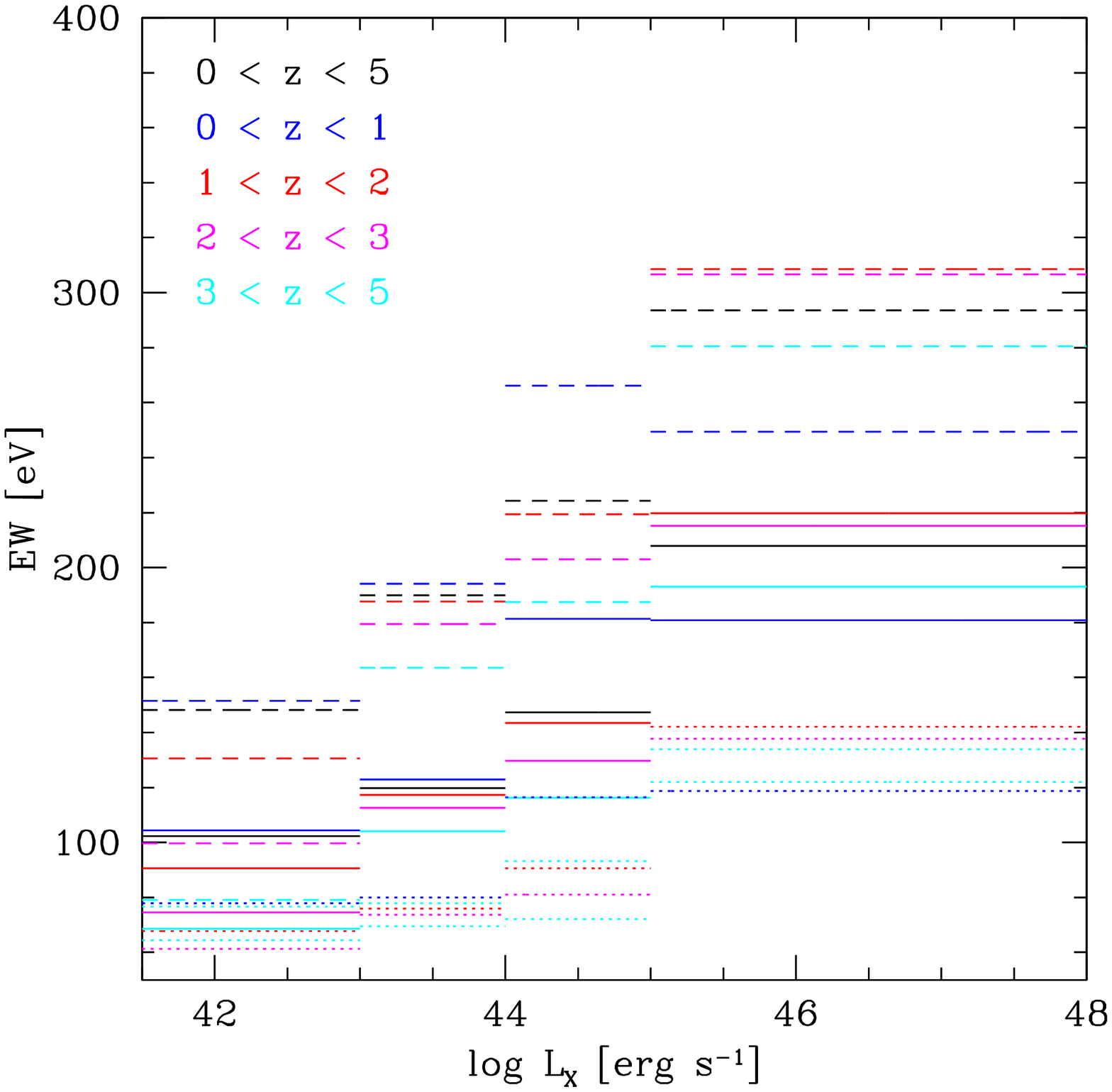}
\caption{\fe\ EWs measured from AGN spectra integrated over
  different ranges of $z$ and $L_{X}$ (Eqn.~\ref{eq:cxrb}). The colors denote the different redshift ranges
  the integrals were performed over with black indicating the result
  from the entire $z=0$ to $5$ interval. The solid lines show the
  results for a reflection fraction of unity (i.e., $R=1$), while the
  dashed and dotted lines plot the EWs if all AGNs have $R=2$ or
  $0.5$, respectively.}
\label{fig:ews}
\end{figure}

\begin{figure}
\includegraphics[angle=-90,width=0.95\textwidth]{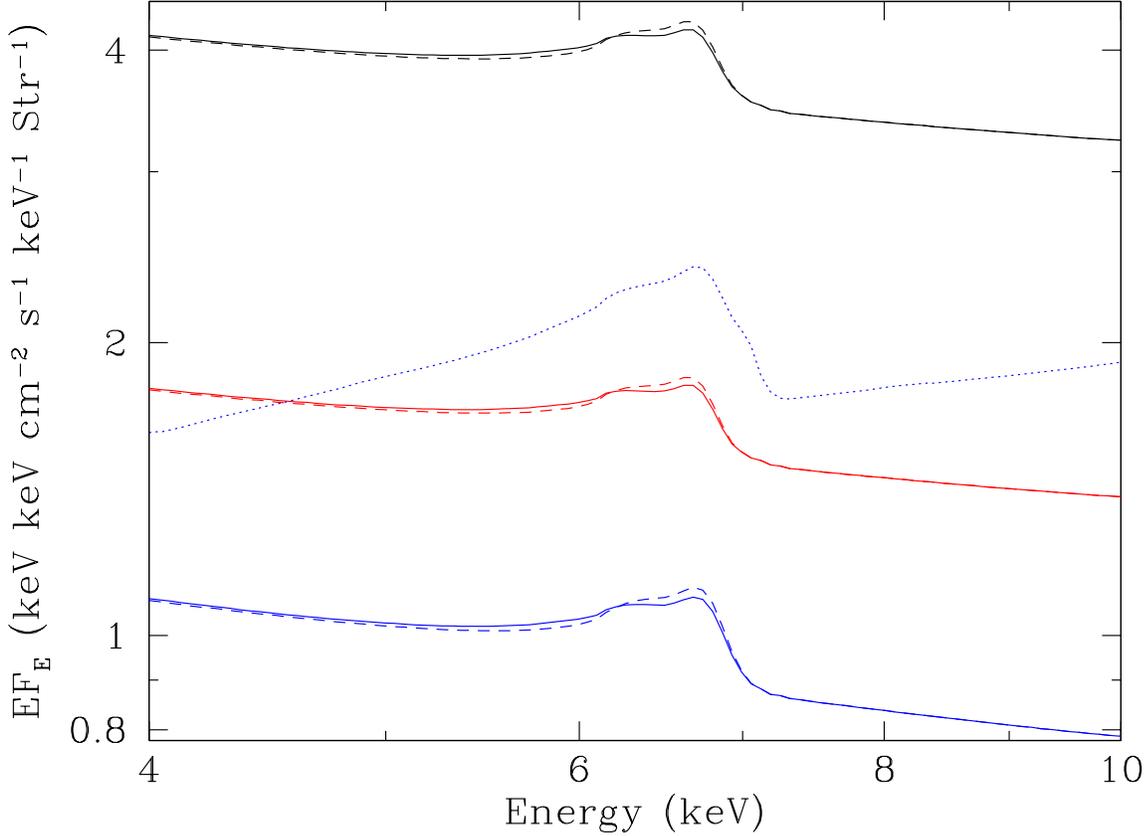}
\caption{Examples of the integrated $R=1$ spectra from which the \fe\ EWs
  were measured. The colors correspond to the same redshift ranges as
  in Fig.~\ref{fig:ews}. The solid lines show the spectra for the $44
  < \log L_{X} \leq 45$ bin assuming the emission line extends down to
  the ISCO of a rapidly spinning black hole. The dashed lines plot the
  same spectra, but now assuming that the ISCO is $6$~$r_g$ for all
  AGNs. Both the solid and dashed
  lines are the integrated spectra including only type 1 AGNs (i.e.,
  those with $\log N_{\mathrm{H}} < 22$). The dotted blue line plots
  the integrated type 2 spectrum for AGNs with $44 < \log L_{X} \leq
  45$ and $z \leq 1$. The lines in this spectrum also assume emission
  down the ISCO of a rapidly spinning black hole. In our model the EW of the broad \fe\ line is
  independent of nuclear obscuration.}
\label{fig:lines}
\end{figure}

Measurements of the spin distribution of SMBHs will have profound consequences on our
understanding of galaxy formation and evolution. The only direct way
to measure SMBH spins is via the broad \fe\ line in AGNs, and attempts
have been made to measure the \fe\ profile
in the spectrum of AGNs that have been integrated over a range of $z$
and $\log L_{X}$. The accuracy and precision of these measurements will
be maximized if the spectral fitting can be performed on \fe\ lines
with the largest possible EW. Therefore, this letter presented
predictions of the integrated relativistic \fe\ EW as functions of $z$ and $\log L_{X}$. Fig.~\ref{fig:ews} showed that the integrated broad \fe\ EW will
be significantly less than 300~eV for all ranges of $z$ and $\log
L_{X}$. In fact, if $R=1$ for most AGNs, the integrated broad component will have
an EW between 100 and 200~eV. Taking into account the relative space
density of AGNs, it is recommended that the search for integrated
broad \fe\ lines be confined to AGNs with luminosities in the range
$44 \leq \log L_{X} < 45$. It is further recommended that these
experiments be limited to type 1 AGNs.

Stepping back from the quantitative results, this letter has shown
that measuring the spin evolution of SMBHs can be done with the broad
\fe\ line. In fact, \ixo\ will be perfectly placed to obtain this
goal, as its sensitivity will allow it to obtain the high-quality
spectra necessary to measure the broad \fe\ lines. \ixo\ spectral
surveys of target lists provided by the \chandra\ deep fields will be a
very powerful tool to trace the spin of SMBHs up to and beyond $z=1$.
  
\acknowledgments
The referee is acknowledged for a detailed and useful report that
greatly improved the paper. 

{}

\clearpage

\clearpage


\begin{thebibliography}{}
\bibitem[\protect\citeauthoryear{Arnaud}{1996}]{arn96} Arnaud, K.A.,
  1996, Astronomical Data Analysis Software and Systems V, eds. Jacoby
  G. and Barnes J., p17, ASP Conf. Series Vol 101
\bibitem[\protect\citeauthoryear{Ballantyne}{2010}]{bal10}
  Ballantyne, D.R., 2010, \apj, 708, L1
\bibitem[\protect\citeauthoryear{Ballantyne \etal}{2001}]{bal01}
  Ballantyne, D.R., Ross, R.R. \& Fabian, A.C., 2001, \mnras, 327, 10
\bibitem[\protect\citeauthoryear{Ballantyne \etal}{2002}]{bfr02}
  Ballantyne, D.R., Fabian, A.C. \& Ross, R.R., 2002, \mnras, 329, L67
\bibitem[\protect\citeauthoryear{Bardeen \etal}{1972}]{bpt72} Bardeen,
  J.M., Press, W.H. \& Teukolsky, S.A., 1972, \apj, 178, 347
\bibitem[\protect\citeauthoryear{Berti \& Volonteri}{2008}]{bv08}
  Berti, E. \& Volonteri, M., 2008, \apj, 684, 822
\bibitem[\protect\citeauthoryear{Bolton}{1972}]{bol72} Bolton, C.T.,
  1972, \nat, 235, 271
\bibitem[\protect\citeauthoryear{Brenneman \& Reynolds}{2006}]{br06}
  Brenneman, L.W. \& Reynolds, C.S., 2006, \apj, 652, 1028
\bibitem[\protect\citeauthoryear{Brusa \etal}{2005}]{bgc05} Brusa, M.,
  Hilli, R. \& Comastri, A., 2005, \apj, 621, L5
\bibitem[\protect\citeauthoryear{Corral \etal}{2008}]{cor08} Corral,
  A. \etal, 2008, \aap, 492, 71 
\bibitem[\protect\citeauthoryear{Cowley \etal}{1983}]{cow83} Cowley,
  A.P., Crampton, D., Hutchings, J.B., Remillard, R. \& Penfold, J.E.,
  1983, \apj, 272, 118
\bibitem[\protect\citeauthoryear{Dai \etal}{2010}]{dai10} Dai, X.,
  Kochanek, C.S., Chartas, G., Koz\l owski, S., Morgan, C.W., Garmire,
  G. \& Agol, E., 2010, \apj, 709, 278
\bibitem[\protect\citeauthoryear{Draper \& Ballantyne}{2009}]{db09}
  Draper, A.R. \& Ballantyne, D.R., 2009, \apj, 707, 778
\bibitem[\protect\citeauthoryear{Fabian \etal}{1989}]{fab89} Fabian,
  A.C., Rees, M.J., Stella, L. \& White, N.E., 1989, \mnras, 238, 729
\bibitem[\protect\citeauthoryear{Gandhi \& Fabian}{2003}]{gf03}
  Gandhi, P. \& Fabian, A.C., 2003, \mnras, 339, 1095
\bibitem[\protect\citeauthoryear{Gandhi \etal}{2007}]{gan07} Gandhi,
  P., Fabian, A.C., Suebsuwong, T., Malzac, J., Miniutti, G., Wilman,
  R.J., 2007, \mnras, 382, 1005
\bibitem[\protect\citeauthoryear{Ghez \etal}{2008}]{ghez08} Ghez, A.,
  \etal, 2008, \apj, 689, 1044
\bibitem[\protect\citeauthoryear{Gilli \etal}{1999}]{gil99} Gilli, R.,
  Comastri, A., Brunetti, G. \& Setti, G., 1999, \na, 4, 45
\bibitem[\protect\citeauthoryear{Inoue \etal}{2007}]{ith07} Inoue, H.,
  Terashima, Y., \& Ho, L.C., 2007, \apj, 662, 860
\bibitem[\protect\citeauthoryear{Iwasawa \etal}{2004}]{iwa04} Iwasawa,
  K., Miniutti, G. \& Fabian, A.C., 2004, \mnras, 355, 1073
\bibitem[\protect\citeauthoryear{King \& Pringle}{2006}]{kp06} King,
  A.R. \& Pringle, J.E., 2006, \mnras, 373, L90
\bibitem[\protect\citeauthoryear{Labita \etal}{2009}]{lab09} Labita,
  M., Decarli, R., Treves, A. \& Falomo, R., 2009, \mnras, 399, 2099
\bibitem[\protect\citeauthoryear{Laor}{1991}]{lao91} Laor, A., 1991,
  \apj, 376, 90
\bibitem[\protect\citeauthoryear{Magorrian \etal}{1998}]{mag98}
  Magorrian, J. \etal, 1998, \aj, 115, 2285
\bibitem[\protect\citeauthoryear{McClintock \& Remillard}{2006}]{mr06}
  McClintock, J.E. \& Remillard, R.A., 2006, in Compact Stellar X-ray
  Sources, ed. W.H.G. Lewin \& M. van der Klis,
  pp. 157-214. Cambridge: Cambridge Univ
\bibitem[\protect\citeauthoryear{Miniutti \etal}{2009}]{min09}
  Miniutti, G., Panessa, F., de Rosa, A., Fabian, A.C., Malizia, A.,
  Molina, M., Miller, J.M. \& Vaughan, S., 2009, \mnras, 398, 255
\bibitem[\protect\citeauthoryear{Nandra \etal}{2007}]{nan07} Nandra,
  K., O'Neill, P.M., George, I.M., \& Reeves, J.N. 2007, \mnras, 382,
  194
\bibitem[\protect\citeauthoryear{Narayan \& Yi}{1995}]{ny95} Narayan,
  R. \& Yi, I., 1995, \apj, 452, 710
\bibitem[\protect\citeauthoryear{Netzer}{2009}]{net09} Netzer, H.,
  2009, \apj, 695, 793
\bibitem[\protect\citeauthoryear{Netzer \& Trakhtenbrot}{2007}]{nt07}
  Netzer, H. \& Trakhtenbrot, B., 2007, \apj, 654, 754
\bibitem[\protect\citeauthoryear{Peterson \etal}{2004}]{peter04}
  Peterson, B.M., \etal, 2004, \apj, 613, 683
\bibitem[\protect\citeauthoryear{Reynolds \& Fabian}{2008}]{rf08}
  Reynolds, C.S. \& Fabian, A.C., 2008, \apj, 675, 1048
\bibitem[\protect\citeauthoryear{Risaliti \etal}{2009}]{rye09}
  Risaliti, G., Young, M. \& Elvis, M., 2009, \apj, 700, L6
\bibitem[\protect\citeauthoryear{Ross \& Fabian}{1993}]{rf93} Ross,
  R.R. \& Fabian, A.C., 1993, \mnras, 261, 74
\bibitem[\protect\citeauthoryear{Schmoll \etal}{2009}]{sch09} Schmoll,
  S., Miller, J.M., Volonteri, M., Cackett, E., Reynolds, C.S.,
  Fabian, A.C., Brenneman, L.W., Miniutti, G. \& Gallo, L.C., 2009,
  \apj, 703, 2171
\bibitem[\protect\citeauthoryear{Spergel \etal}{2003}]{spe03} Spergel,
  D.N., \etal, 2003, \apjs, 148, 175
\bibitem[\protect\citeauthoryear{Streblyanska \etal}{2005}]{streb05}
  Streblyanska, A., Hasinger, G., Finoguenov, A., Barcons, X., Mateos,
  S. \& Fabian, A.C., 2005, \aap, 432, 395
\bibitem[\protect\citeauthoryear{Ueda \etal}{2003}]{ueda03} Ueda, Y.,
  Akiyama, M., Ohta, K. \& Miyaji, T., 2003, \apj, 598, 886
\bibitem[\protect\citeauthoryear{Volonteri \etal}{2005}]{v05}
  Volonteri, M., Madau, P., Quateart, E. \& Rees, M.J., 2005, \apj,
  620, 69
\end{thebibliography}
\end{document}